\documentclass{ieeeaccess}
\usepackage{cite}
\usepackage{amsmath,amssymb,amsfonts}
\usepackage{algorithmic}
\usepackage[braket, qm]{qcircuit}
\usepackage{graphicx}
\usepackage{textcomp}
\usepackage{url}
\usepackage{multirow}

\usepackage{bm}
\makeatletter
\AtBeginDocument{\DeclareMathVersion{bold}
\SetSymbolFont{operators}{bold}{T1}{times}{b}{n}
\SetSymbolFont{NewLetters}{bold}{T1}{times}{b}{it}
\SetMathAlphabet{\mathrm}{bold}{T1}{times}{b}{n}
\SetMathAlphabet{\mathit}{bold}{T1}{times}{b}{it}
\SetMathAlphabet{\mathbf}{bold}{T1}{times}{b}{n}
\SetMathAlphabet{\mathtt}{bold}{OT1}{pcr}{b}{n}
\SetSymbolFont{symbols}{bold}{OMS}{cmsy}{b}{n}
\renewcommand\boldmath{\@nomath\boldmath\mathversion{bold}}}
\makeatother

\def\BibTeX{{\rm B\kern-.05em{\sc i\kern-.025em b}\kern-.08em
    T\kern-.1667em\lower.7ex\hbox{E}\kern-.125emX}}

\begin{document}
\history{Date of publication xxxx 00, 0000, date of current version xxxx 00, 0000.}
\doi{10.1109/ACCESS.2024.0429000}

\title{Quantitative Evaluation of Quantum/Classical Neural Network Using a Game Solver Metric}

\author{\uppercase{Suzukaze Kamei}\authorrefmark{1}, 
\uppercase{Hideaki Kawaguchi}\authorrefmark{2},
\uppercase{Shin Nishio}\authorrefmark{2,3},
and \uppercase{Takahiko Satoh}\authorrefmark{1}}

\address[1]{Faculty of Science and Technology, Keio University, Yokohama, Kanagawa 223-8522 Japan}
\address[2]{Graduate School of Science and Technology, Keio University, Yokohama, Kanagawa 223-8522 Japan}
\address[3]{Department of Physics \& Astronomy, University College London, London, WC1E 6BT, United Kingdom}
\tfootnote{This work was supported by JST Moonshot R\&D Grant Number JPMJMS226C. SK and TS are also supported by JST COI-NEXT Grant Number JPMJPF2221 and MEXT KAKENHI Grant Number 22K1978. SN is also supported by JSPS KAKENHI Grant Number JP22KJ1436 and JSPS Overseas Research Fellowships.}

\markboth
{Kamei \headeretal: Quantitative Evaluation of Quantum/Classical Neural Network Using a Game Solver Metric}
{Kamei \headeretal: Quantitative Evaluation of Quantum/Classical Neural Network Using a Game Solver Metric}

\corresp{Corresponding author: Suzukaze Kamei (e-mail: suzukaze\_kamei@keio.jp).}

\begin{abstract}
To evaluate the performance of quantum computing systems relative to classical counterparts and explore the potential, we propose a game-solving benchmark based on Elo ratings in the game of tic-tac-toe. We compare classical convolutional neural networks (CCNNs), quantum or quantum convolutional neural networks (QNNs, QCNNs), and hybrid classical-quantum neural networks (Hybrid NNs) by assessing their performance based on round-robin matches. Our results show that the Hybrid NNs engines achieve Elo ratings comparable to those of CCNNs engines, while the quantum engines underperform under current hardware constraints. Additionally, we implement a QNN integrated with quantum communication and evaluate its performance to quantify the overhead introduced by noisy quantum channels, and the communication overhead was found to be modest. These results demonstrate the viability of using game-based benchmarks for evaluating quantum computing systems and suggest that quantum communication can be incorporated with limited impact on performance, providing a foundation for future hybrid quantum applications.
\end{abstract}

\titlepgskip=-21pt

\maketitle

\section{Introduction}
\label{sec:introduction}

\PARstart{Q}{uantum} computers have brought new possibilities to many previously difficult problems for classical computers to solve~\cite{Shor_1997, grover1996fast}. Furthermore, the emergence of an algorithm that allows many existing algorithms to be viewed from a unified perspective is also accelerating the development of new quantum algorithms~\cite{PRXQuantum.2.040203, 10.1145/3549524}. In recent years, the development of quantum computers has made remarkable progress. Noisy Intermediate-Scale Quantum (NISQ) computers exceeding 1,000 qubits have emerged, and research and development for  Fault-Tolerant Quantum Computers (FTQC) has become very active~\cite{yoder2025tourgrossmodularquantum}.
With further advancements in both algorithms and hardware, the potential for applying quantum advantage to real-world problems is increasingly becoming a reality.

Current benchmarks for quantum computers, such as random circuit sampling, demonstrate quantum advantages in limited tasks~\cite{morvan2024phase}. However, evidence supporting their superiority in solving real-world problems remains inconclusive~\cite{kim2023evidence}. Furthermore, the absence of common and standardized metrics for computational performance presents a significant challenge in evaluating the potential societal and practical impact of future quantum computing technologies. Therefore, it is necessary to measure the performance of quantum systems from multiple angles using various benchmarking methods, such as Quantum Volume~\cite{cross2019validating} and Circuit Mirroring~\cite{proctor2022measuring}, the latter being a flexible technique capable of generating customized, efficiently verifiable benchmarks based on any circuit structure, which can even transform Quantum Volume into a fully scalable measure.

Classical game AIs, such as AlphaZero~\cite{doi:10.1126/science.aar6404}, have been highly successful and have significantly outperformed human abilities in perfect-information zero-sum games. While quantum game AIs have been proposed for solving single-player puzzle games using search algorithms~\cite{IBMQchallenge2020}, algorithms that can adapt to dynamically changing board information in competitive games remain unexplored.

By utilizing a unified metric, it becomes possible to compare the performance of quantum and classical systems directly. Being able to compare these systems in a unified environment allows us to evaluate the progress of quantum systems not only from a quantum perspective but also from a classical one. This has the potential to serve as a key benchmark for demonstrating quantum advantage.

\begin{figure*}[htbp]
\begin{center}
\includegraphics[width=0.9\linewidth]{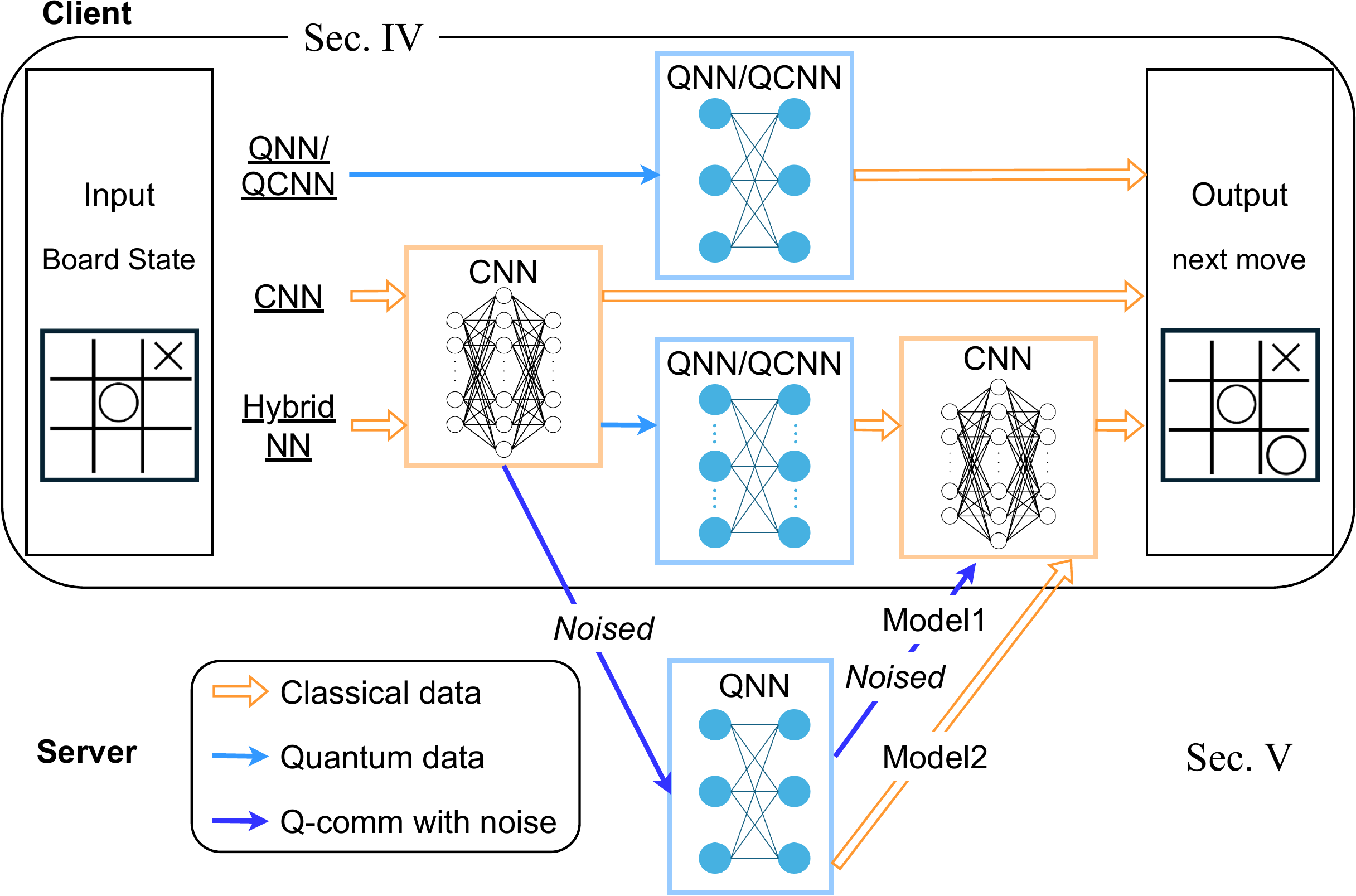}
\caption{
The game solver initially runs on the client, receiving the board state and selecting the next move using a classical, quantum, or hybrid neural network.
In the Quantum Internet setup, a noisy quantum communication (Q-comm) channel connects the client and server. The client processes the input with a classical neural network, encodes the output into qubits, and sends them through the quantum channel. The server then applies a quantum neural network and returns either classical measurement results or unmeasured qubits. The client uses the received data to determine the next move.
}
\label{fig:overview}
\end{center}
\end{figure*}
In this research, we propose a novel approach for evaluating the performance of quantum and classical computers through competitive board games.
An overview of the proposed game-solving framework—including client-side inference, quantum communication, and server-side processing—is illustrated in Fig.~\ref{fig:overview}.
Game-playing AI provides a suitable platform for comprehensive performance assessment, as it involves data structure manipulation, decision-making, and real-time optimization.

The objective of this research is to quantitatively assess the capabilities of quantum computers using competitive games. To this end, we adopt the following methodology: 
\begin{enumerate} 
\item Develop a game engine incorporating classical and quantum neural networks. 
\item Design three policy types—classical, quantum, and classical-quantum hybrid—and compare their performance. \item Evaluate each engine using Elo ratings to provide a unified performance metric. 
\end{enumerate}

\begin{figure}[htbp]
\begin{center}
\includegraphics[width=0.35\linewidth]{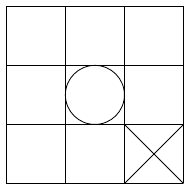}
\caption{Tic-Tac-Toe game board consisting of 9 squares. 2 players place O and X alternately in the squares. The first player to place O or X in a vertical, horizontal, or diagonal line wins the game.
If neither of them is aligned, the game ends in a tie.}
\label{fig: tictactoe}
\end{center}
\end{figure}

While this game is extremely simple and cannot concretely clarify areas where quantum computers outperform classical methods, the design that can adapt to more complex games is important for this unexplored topic. 

Furthermore, we evaluate the impact of connecting quantum devices with varying performance levels, in anticipation of future quantum communication applications.
Just as classical high-performance computing is widely accessed via the cloud, similar usage scenarios are expected for quantum devices. In this research, we examine the practicality of quantum applications over the quantum Internet (QI) by analyzing the communication overhead caused by noise, assuming a client with limited quantum resources and a server with abundant ones.
While the detailed QI protocols involving entanglement are beyond our scope, we construct and test simplified models under two scenarios, based on the client’s quantum resource availability.

To explore this approach, we first review the machine learning techniques and Elo rating system used in our work (Sec.~\ref{sec_preliminaries}).
We then describe the architecture of the proposed game engines (Sec.~\ref{sec_engines}) and evaluate their performance (Sec.~\ref{sec_eval}).
Next, we introduce the QI model and assess the performance of game engines operating within this setting (Sec.~\ref{sec_qimodel}).
Finally, we give a summary and discuss future research directions (Sec.~\ref{sec_conclusion}).

\section{Preliminaries}
\label{sec_preliminaries}

\subsection{reinforcement learning}
\label{sub_reinforce}

Reinforcement learning~\cite{sutton1998reinforcement ,watkins1989learning, mnih2013playing} is a method of learning to make the optimal choice in a particular environment.
Unlike supervised and unsupervised learning, it aims to learn through the interaction between the environment and the agent and maximize the reward so that the agent can cope with unknown domains.

\begin{figure}[htbp] 
\begin{center}
\includegraphics[width=0.8\linewidth]{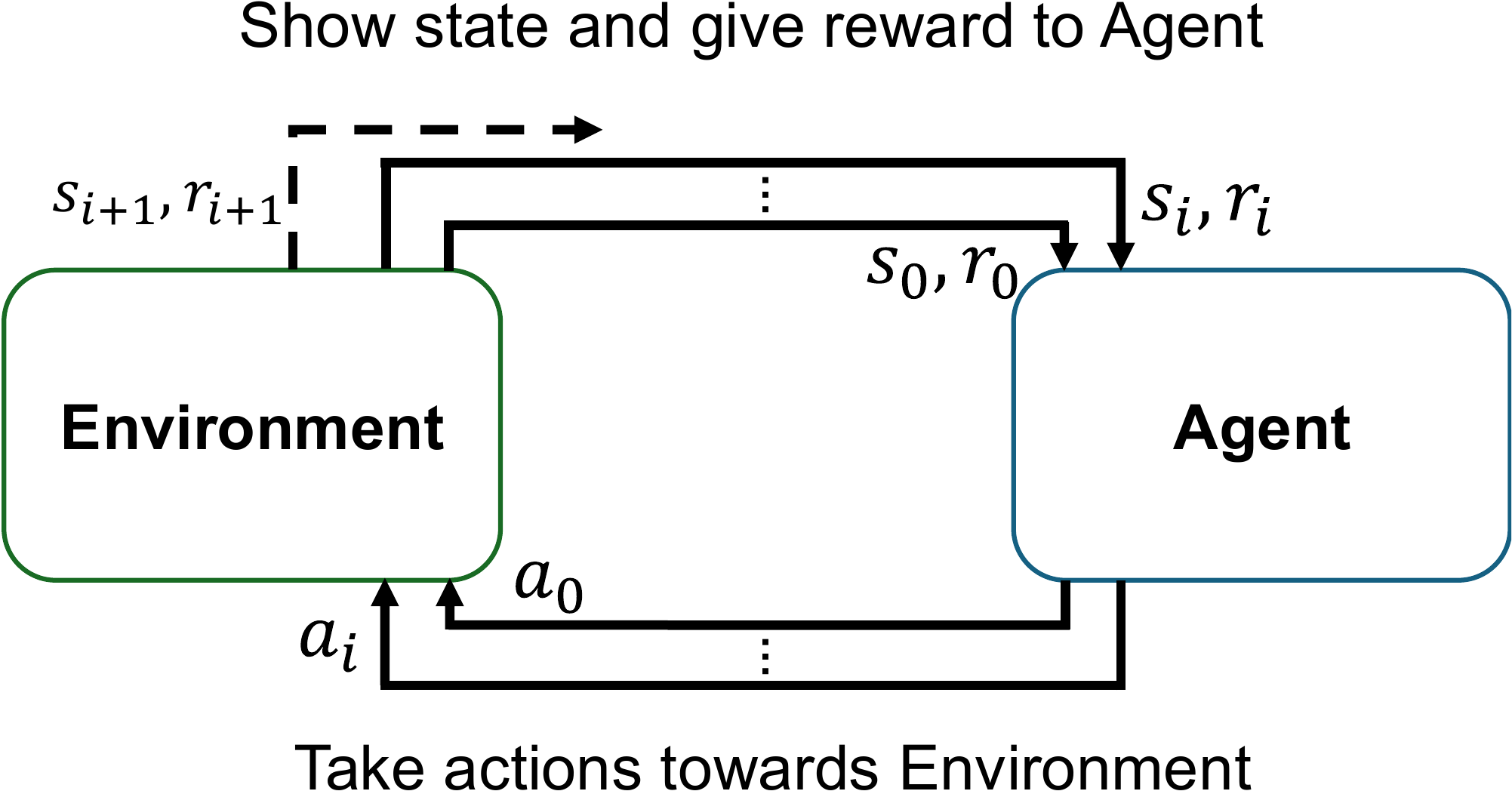}
\caption{
Schematic diagram of agent–environment interaction in reinforcement learning.
At each time step $i$, the environment provides the current board state $s_i$ and reward $r_i$ to the agent. 
The agent then selects an action $a_i$ based on this input. 
The environment updates the board accordingly and returns the next state $s_{i+1}$ and reward $r_{i+1}$. 
This cycle is repeated throughout the game.
}
\label{fig: reinforce}
\end{center}
\end{figure}

For example, in the case of tic-tac-toe, the environment corresponds to the game board, and the agent is the game engine incorporating the neural network. Learning proceeds based on the board state, the reward received, and the next move determined by the engine (Fig.~\ref{fig: reinforce}) .

There is a strong connection between reinforcement learning and game AI, especially in games such as chess and shogi~\cite{doi:10.1126/science.aar6404, Schrittwieser_2020}. Competitive games are well-suited for reinforcement learning because outcomes (win/loss) are clearly defined and player performance can be quantitatively evaluated. As reinforcement learning algorithms have evolved, game AI has improved in parallel.

Q-learning~\cite{watkins1992q} is one of the foundational algorithms in reinforcement learning. In this research, we developed and evaluated a game engine based on Q-learning. This section provides a brief overview of the algorithm.

Reinforcement learning aims to maximize the expected return from a given strategy. When the environment follows a Markov decision process, the expected value can be divided into two types: a \textit{state-value function}, which estimates the value of a state, and an \textit{action-value function}, which estimates the value of taking a specific action in a given state.

Q-learning enables efficient reinforcement learning by iteratively updating the action-value function. The update rule is defined as follows:
\begin{multline}\label{math:Q}
Q(s_{i}, a_{i}) \leftarrow Q(s_{i}, a_{i}) \\
+ \alpha \Bigl[ 
    r_{i+1} + \gamma \max_{a_{i+1}} Q(s_{i+1}, a_{i+1}) 
    - Q(s_{i}, a_{i})
\Bigr].
\end{multline}

Here, $Q(s_i, a_i)$ represents the estimated return of taking action $a_i$ in state $s_i$.  
$\alpha$ is the learning rate, and $\gamma$ is the discount factor, which determines the importance of future rewards.  
$r_{i+1}$ is the reward received after performing action $a_i$, and $s_{i+1}$ is the resulting next state.  
The update adjusts the current estimate toward the target value, which consists of the immediate reward plus the discounted maximum future value.

\subsection{Quantum Machine Learning}
\label{sub_qml}
Quantum machine learning~\cite{biamonte2017quantum, schuld2015introduction} has emerged as a promising application of quantum computing. It aims to develop learning algorithms that leverage quantum properties to outperform classical machine learning in terms of computational efficiency and model expressiveness.

In particular, quantum neural networks (QNNs)~\cite{gupta2001quantum, beer2020training} have been proposed as quantum analogues of classical neural networks. When processing classical data, a typical QNN pipeline involves three stages: encoding the data into quantum states, processing it using parameterized quantum circuits (ansatz), and measuring the output to obtain classical values. These values are used to optimize parameters via classical computation.

\begin{figure}[htbp]
\centering
\includegraphics[width=0.5\linewidth]{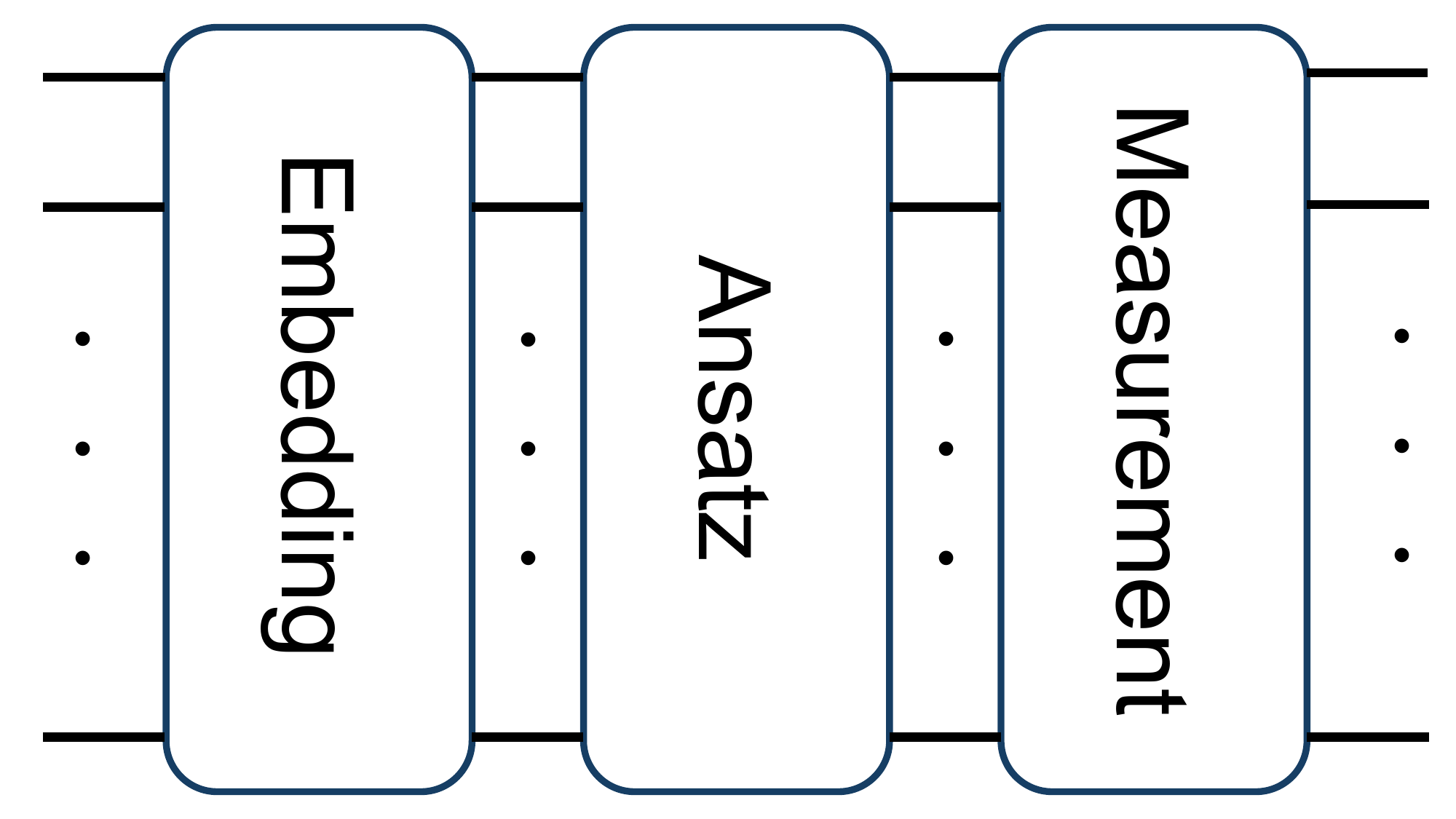}
\caption{Structure of a quantum neural network (QNN). A typical QNN consists of three components: (1) \textit{Embedding} – classical data is encoded into quantum states $\vert \psi_k \rangle$; (2) \textit{Ansatz} – parameterized quantum circuits optimized via classical computation; and (3) \textit{Measurement} – quantum states are measured to produce classical outputs.}
\label{fig: qml}
\end{figure}

\subsubsection*{Quantum Neural Network Architectures}

Quantum machine learning offers a variety of neural network architectures, each designed to address different aspects of performance and scalability.

One such approach is the hybrid neural network (HNN), which combines classical and quantum components. For example, classical neural networks (CNNs) may be placed before or after a quantum layer, or quantum circuits may be used to perform certain computations such as gradient descent~\cite{kerenidis2020quantum, stokes2020quantum}. These hybrid models aim to exploit the strengths of both paradigms, making them suitable for near-term quantum devices.

Another notable architecture is the quantum convolutional neural network (QCNN)~\cite{cong2019quantum}, which is inspired by classical convolutional networks but designed to operate on quantum states. Compared to general QNNs, QCNNs can learn more efficiently with fewer parameters, as the number of parameters scales logarithmically with the number of qubits.  Moreover, they have been shown to mitigate the \textit{barren plateau} problem~\cite{pesah2021absence}, in which gradients vanish during training, hindering effective optimization.

The QCNN ansatz consists of alternating convolution and pooling layers (Fig.~\ref{fig: embed_and_ansatz}). In the convolutional layers, entanglement between qubits is used to extract local features. While the pooling layers in classical CNNs aggregate information, its quantum counterparts achieve dimensionality reduction by concentrating information onto a single qubit, which is then discarded. Repeating these layers enables hierarchical feature extraction and supports scalable learning.

These architectures demonstrate how quantum machine learning can be structured to balance expressivity and trainability. Hybrid models are particularly promising in the current era of NISQ devices, while QCNNs offer a scalable framework for leveraging quantum entanglement and feature abstraction in learning tasks.

\subsection{elo rating}
\label{sub_elo}

Elo rating~\cite{elo1978, albers2001elo, HVATTUM2010460} is a widely used system for quantifying the strength of individual players in competitive games, such as chess. It enables an approximate comparison of player skill levels, regardless of game order or the number of opponents.

In our setting, we define the Elo rating of player A as $R_A$ and that of player B as $R_B$. When A and B compete, the probability that A wins is given by:
\begin{align}
W_{AB} = \frac{1}{10^{(R_B - R_A)/400} + 1}
\label{eq:elo_win_prob}
\end{align}
Using this probability, the rating of player A is updated as:
\begin{align}
R'_A = R_A + K (N_{\text{wins}} - N_{\text{games}} \cdot W_{AB}),
\label{eq:elo_update}
\end{align}
where $N_{\text{wins}}$ is the number of games A won, $N_{\text{games}}$ is the total number of games played (excluding draws), and $K$ is a scaling factor that controls the update sensitivity.

\begin{figure}[htbp]
\begin{center}
\includegraphics[width=0.7\linewidth]{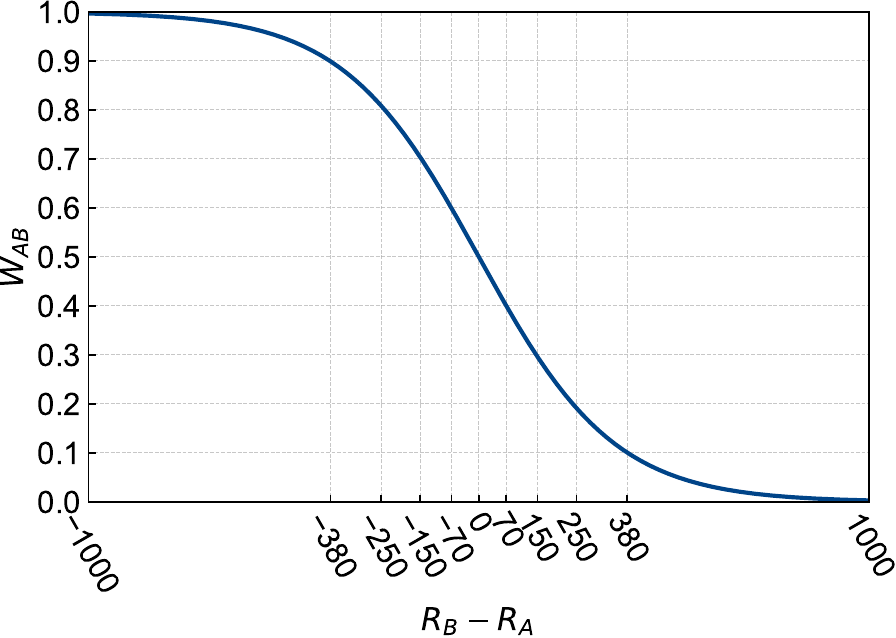}
\caption{Relationship between rating difference and winning rate.}
\label{fig: elo}
\end{center}
\end{figure}

Figure~\ref{fig: elo} shows the relationship between the rating difference $(R_B - R_A)$ and the expected winning probability $W_{AB}$, as defined in Eq.~\eqref{eq:elo_win_prob}. From the curve, we observe that a rating difference of approximately 70 corresponds to a winning probability of about 60\%.

\subsection{Quantum Machine Learning with Client-Server model}
\label{sub_q-comm}
\subsubsection{Overview of the Client-Server Model}
Quantum computers require large-scale infrastructure, so they are typically owned by major vendors rather than individual users. In this situation, the client sends some classical or quantum information to the server owned by the vendors, and then the vendor executes the costly part of the computation.
This type of computational setup is referred to as the client-server model, and it is regarded as an application of the quantum Internet. The client-server model not only enables physically distributed computing systems but also offers a variety of additional advantages when quantum machine learning tasks use this model\cite{araujo2024quantum, song2024quantum}. One can consider a scenario where the client-server model is adopted in the following manner: the client encodes classical data into quantum states and sends them to the server; the server calculate ansatz to perform the necessary quantum operations; then, the resulting quantum states are returned to the client, who performs the measurements and processes the outcomes.

\subsubsection{Advantages of the Model}
One of the most significant benefits of this model is data privacy. Since the raw classical data remains on the client side and is embedded into quantum states locally, the server, which may be untrusted or a third-party quantum cloud provider, never gains access to the original data. This structure allows secure delegation of quantum resources while ensuring that sensitive input information remains protected.

Another advantage is greater flexibility in classical optimization. The client retains full responsibility for processing measurement results, computing loss functions, and updating parameters during optimization. This separation allows integration with a wide range of classical optimization algorithms without needing to expose internal training dynamics to the server. Additionally, because the quantum server performs only the ansatz operations, this model helps reduce quantum circuit depth, which is particularly important when working with near-term quantum devices (NISQ era) that suffer from decoherence and gate errors.

\subsubsection{Roles and Use Cases}
Moreover, this model supports a clear separation of roles between data owners and model providers. The client may represent an entity with proprietary data, while the server hosts a proprietary model or provides quantum computing resources. This separation encourages secure collaboration and scalability. Importantly, the model also opens doors to federated or blind quantum learning, where multiple clients can train using local data and shared quantum resources without exposing their private information to each other or the server. 

\subsubsection{Implications for Noisy Quantum Communication}
Finally, having the client perform measurements locally can lead to greater control over noise and readout errors, avoiding server-side biases or inaccuracies in quantum measurement. Together, these benefits make this client-server model a promising candidate for secure, efficient, and scalable quantum machine learning in real-world applications.

\section{Tic-Tac-Toe Engines}
\label{sec_engines}

In this section, we describe three types of game engines, each utilizing a classical, quantum, or hybrid neural network architecture. All neural networks are trained using classical optimization methods. Quantum circuits are integrated into the PyTorch framework via the \texttt{TorchConnector} provided by Qiskit~\cite{qiskit2024}. We employ Huber Loss~\cite{huber1992robust} as the loss function and the Adam optimizer~\cite{kingma2014adam} for weight updates.

\begin{figure}[ht]
\begin{center}
\includegraphics[width=0.9\linewidth]{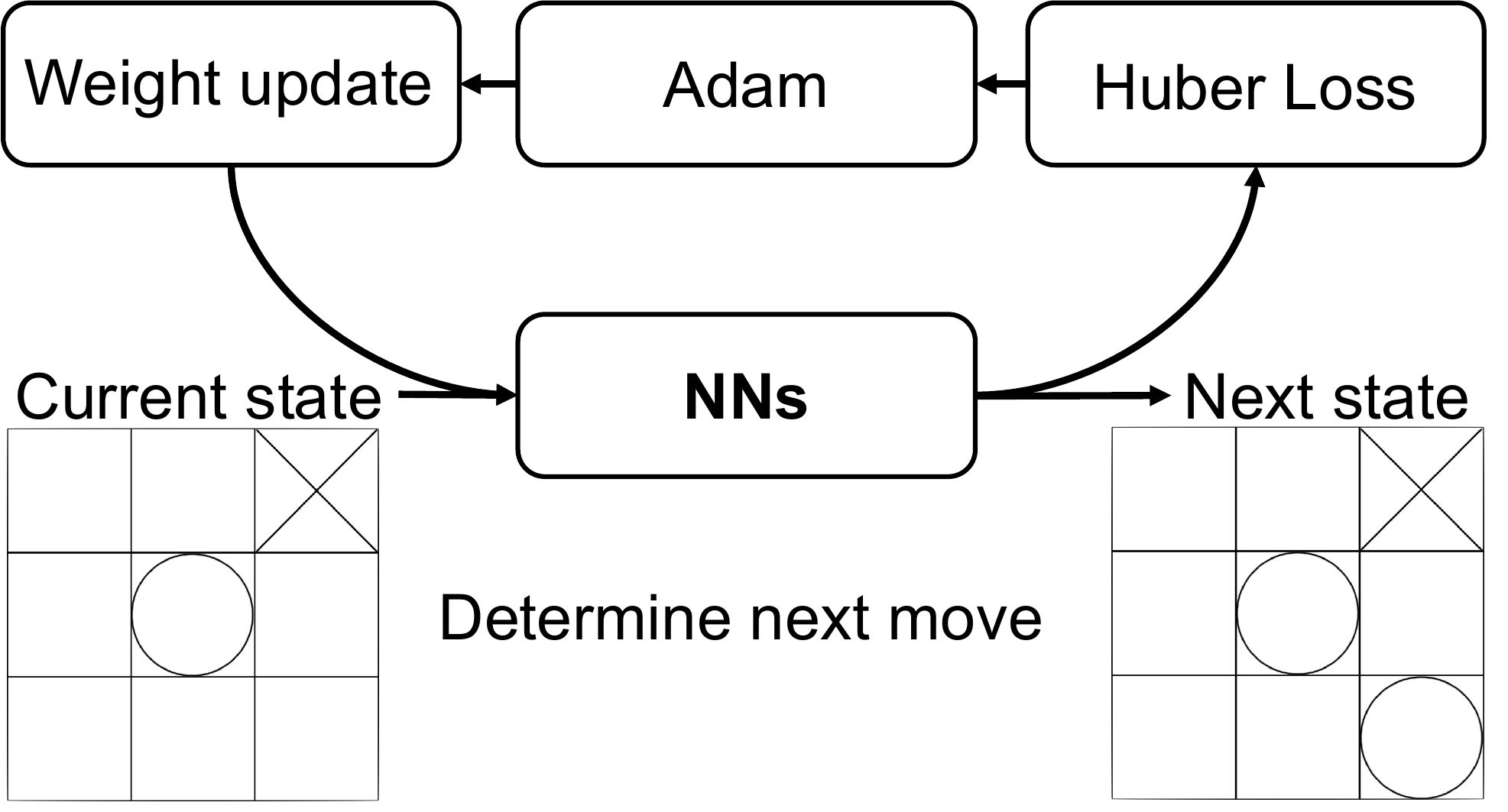}
\caption{Scheme of neural network-based engines. NNs such as QNN (Fig.~\ref{fig: qml}), CNN, and QCNN are integrated into the engine. The input game board is encoded into numerical values and passed through the network to generate an output vector. The engine selects the next move based on the cell with the highest output value. Training is performed using Huber Loss and the Adam optimizer to update network weights.}
\label{fig: scheme}
\end{center}
\end{figure}

The overall structure of the neural network-based engines is illustrated in Fig.~\ref{fig: scheme}. Since neural networks require numerical input and output, the tic-tac-toe board is encoded as a 9-element vector, where O is represented as +1, X as –1, and empty squares as 0. The network outputs a vector of nine values, each corresponding to a board cell, and the cell with the highest value is chosen as the next move.

We adopt a simple reinforcement learning scheme by replacing the conventional Q-table with a neural network model. The Q-learning update rule, described in Eq.~\eqref{math:Q}, is applied after each game, with rewards of +1 for a win, –1 for a loss, and 0 for a draw.

\subsection{Classical neural networks (CNN, CCNN)}

\begin{figure}[ht]
\begin{center}
\includegraphics[width=0.9\linewidth]{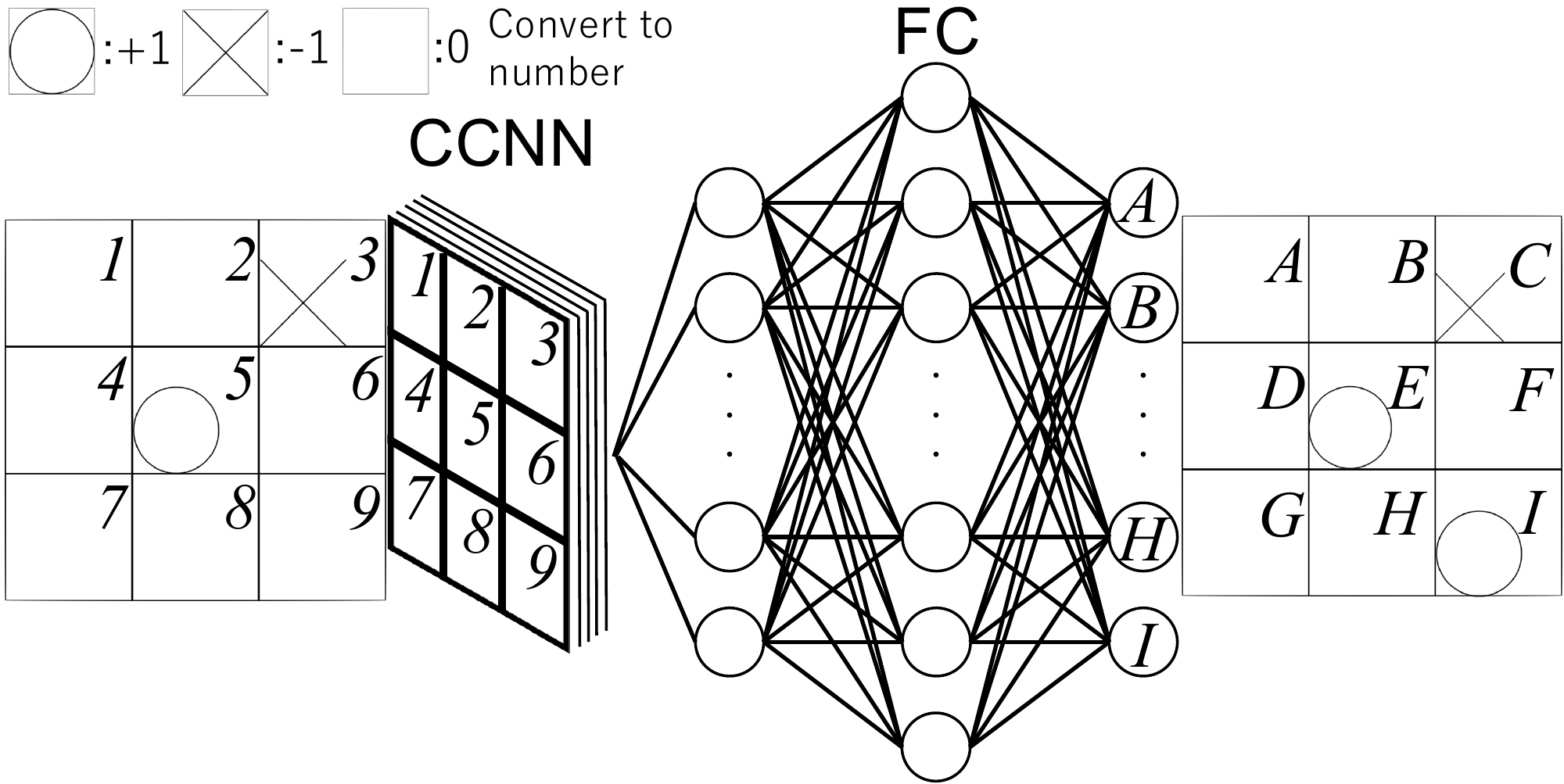}
\caption{Data flow in the classical convolutional neural network (CCNN) engine.  
The tic-tac-toe board is encoded into numerical values and passed through convolutional layers, followed by fully connected (FC) layers.  
The output consists of nine values corresponding to the board cells, and the move with the highest score is selected.}
\label{fig: classical engine}
\end{center}
\end{figure}

The engine is constructed by combining CNNs and CCNNs (Fig.~\ref{fig: classical engine}).
The engine using CCNNs was structured to perform convolution on the input board, and then to obtain output using CNNs.
Two variants, referred to as "Stronger" and "Weaker," differ in the number of layers and parameters. The architectures are summarized in Table~\ref{table: NNArc}.

\subsection{Quantum neural networks (QNN, QCNN)}

The QNN-based engines differ in their choice of embedding circuits (\texttt{ZFeatureMap}, \texttt{ZZFeatureMap}, \texttt{HEE}, \texttt{TPE}), and ansatz circuits (\texttt{RealAmplitudes}, \texttt{EfficientSU2}). All models use 9 qubits, matching the size of board.

\begin{figure}[ht]
\begin{center}
\includegraphics[width=0.9\linewidth]{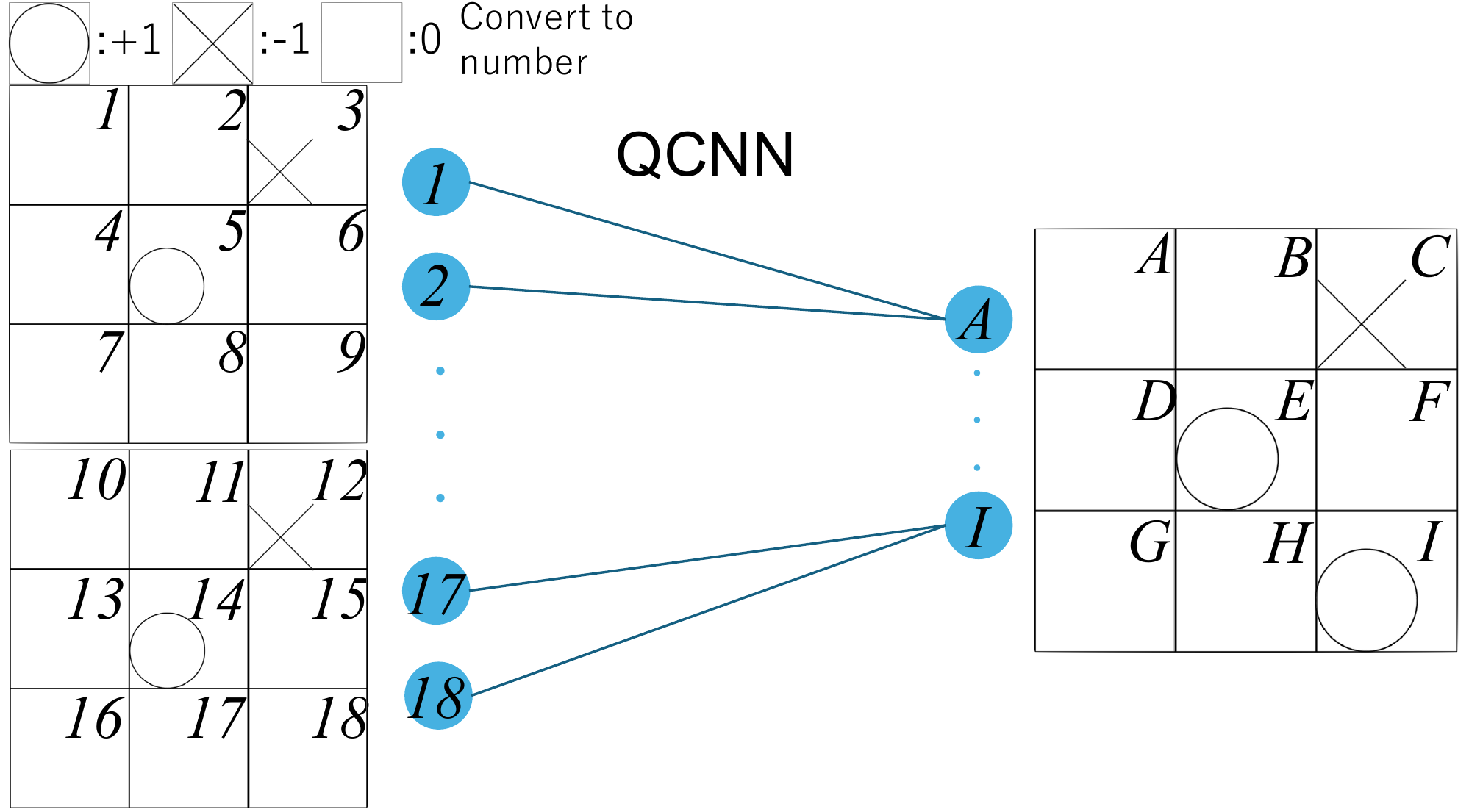}
\caption{Data flow in the quantum convolutional neural network (QCNN) engine.  
The board is encoded as a 9-dimensional vector and duplicated to match the 18-qubit input requirement.  
After passing through the QCNN layer, the output is mapped to nine values representing board cells. The move with the highest score is selected.}
\label{fig: qcnn engine}
\end{center}
\end{figure}

The QCNN-based engines (Fig.~\ref{fig: qcnn engine}) differ in their choice of embedding circuits mensioned above, and ansatz circuit is QCNN. All models use 18 qubits, with the 9-cell iput duplicated to match the quantum register size. Details of the quantum circuit designs are provided in Appendix~\ref{app_qc}.

The number of shots in the measurement is set to 1024, and half of each qubit is measured in the Z basis, and the expected value is taken as the output.

\subsection{Hybrid neural networks (HNN)}
\begin{figure}[ht]
\begin{center}
\includegraphics[width=0.9\linewidth]{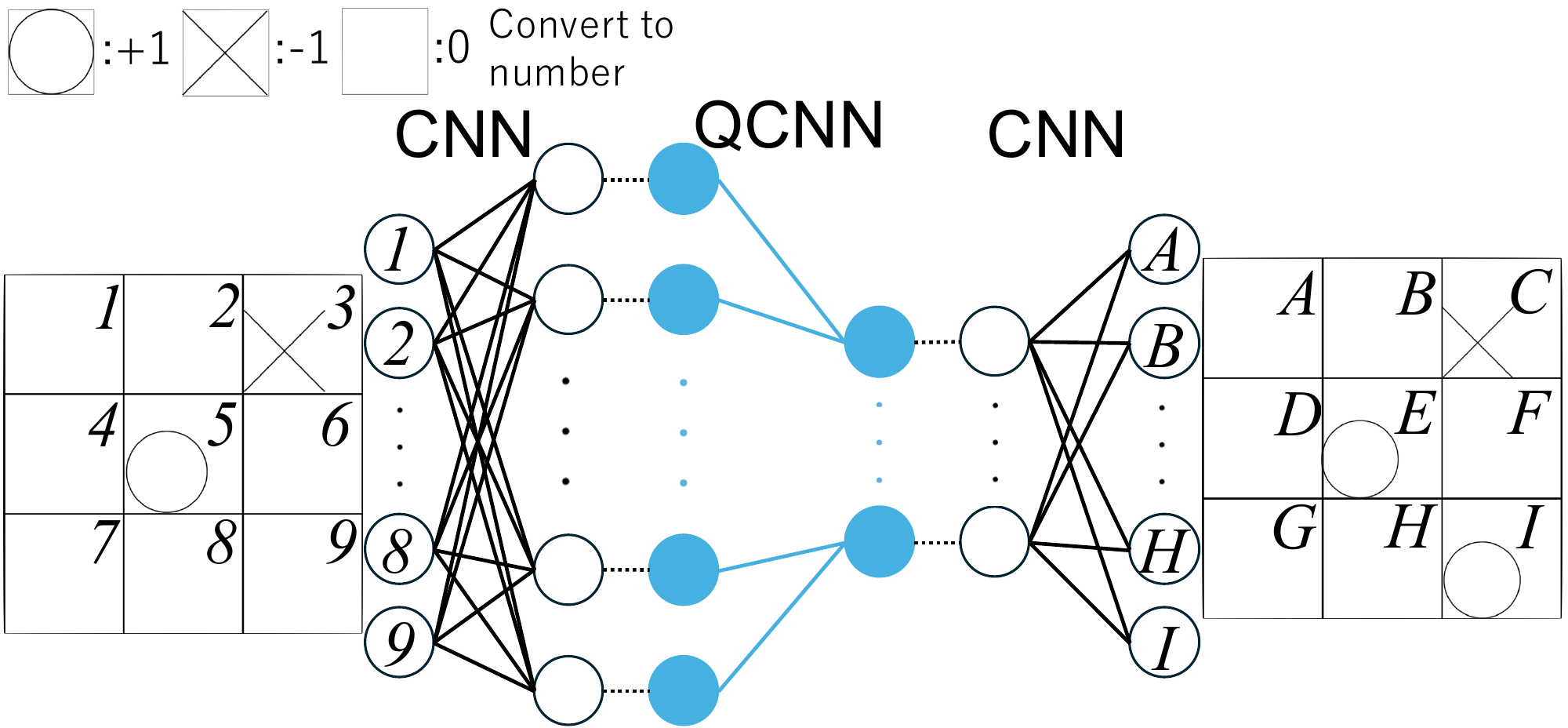}
\caption{Data flow in the hybrid classical-quantum neural network (HNN) engine.  
Encoded board data is processed through an input CNN layer, followed by a quantum layer (QNN or QCNN), and a final output CNN layer.  
The resulting nine output values are used to determine the next move.}
\label{fig: cqc engine}
\end{center}
\end{figure}

The hybrid engine combines classical and quantum neural network components.  
The input and output layers are implemented using classical convolutional neural networks (CNNs), while the hidden layer consists of a quantum neural network, either a QNN or a QCNN, depending on the configuration (Fig.~\ref{fig: cqc engine}). 
We refer to this architecture as HNN, based on the order in which the classical and quantum neural network components are combined.

To accommodate quantum processing, the input board state is first converted into a 9-dimensional vector using the same encoding as in the classical models (O: +1, X: –1, empty: 0).  
This vector is then expanded to match the number of qubits used in the quantum layer (8 or 16 qubits, depending on the setup), and embedded into quantum states using one of the four encoding methods:
\texttt{ZFeatureMap}, \texttt{ZZFeatureMap}, \texttt{HEE}, or \texttt{TPE}.

Two types of quantum neural network layers are used:
\begin{itemize}
    \item \textbf{QNN}: A generic quantum circuit without convolutional structure.
    \item \textbf{QCNN}: A structured ansatz with convolution and pooling layers, as shown in Fig.~\ref{fig: embed_and_ansatz}.
\end{itemize}

The quantum layer is followed by measurement in the Z-basis and the number of shots in the measurement is set to 1024.  
We consider two measurement strategies:
\begin{itemize}
    \item \textbf{Estimator}: Returns the expectation value of observables.
    \item \textbf{Sampler}: Returns the frequency distribution of measurement outcomes, treated as probabilities.
\end{itemize}
In the QNN-based configurations, two types of parameterized quantum circuits (ansatz) are used: \texttt{EfficientSU2} and \texttt{RealAmplitudes}, both provided by Qiskit.  
In contrast, the QCNN-based models use a fixed QCNN ansatz with a convolution–pooling structure, and do not incorporate these two ansatze types.

The measured values are passed to the output CNN layer, which produces a 9-dimensional vector corresponding to the tic-tac-toe board cells. The cell with the highest value is selected as the next move.  
Training is performed using the same reinforcement learning scheme and optimization methods as described in the previous sections.

We output from the quantum layer in two different methods, Estimator and Sampler as mensioned above.
Estimator outputs the expected value for each observable after the measurement.
Sampler outputs the quasi-probabilities after the measurement.
In these engines we used, qubits are measured in the Z basis.

\section{Evaluation of Engines}
\label{sec_eval}
In this section, we evaluate 54 game engine configurations across classical, quantum, and hybrid neural network architectures.  
Each engine is trained using the reinforcement learning method described in Section~\ref{sec_preliminaries} and evaluated using Elo ratings based on round-robin matches.
\subsection{Overview of Evaluation Settings}
We present a summary of engine types, training settings, and the resulting performance metrics.

The 54 engines are categorized as follows:

\begin{enumerate}
    \item \textbf{Classical CNN engines:} 2 variants with different network depths (see Table~\ref{table: NNArc} (a)).
    \item \textbf{QNN engines:} 8 configurations using different embedding and ansatz circuits (Table~\ref{table: NNArc} (b)).
    \item \textbf{QCNN engines:} 4 configurations using different embedding types (Table~\ref{table: NNArc} (c)).
    \item \textbf{Hybrid NN engines (QNN layer):} 32 variants (2 measurement methods × 2 qubit counts × 8 configurations).
    \item \textbf{Hybrid NN engines (QCNN layer):} 8 variants with 2 qubit settings and 4 embedding types.
\end{enumerate}

\begin{table}[htbp]
\caption{NN Architectures: (a) CNN: Stronger has 3 layers, Weaker has 2, both using \(3 \times 3\) kernels. Stronger also has larger fully connected layers. (b) QNN: A single QNN layer using 9 qubits. (c) QCNN: A single QCNN layer, where the input board is duplicated to match its inputs. (d) HNN: \(N = 8\) or \(16\) for QNN, \(8\) or \(16\) for QCNN. HNNs with a QNN hidden layer support Sampler and Estimator outputs, while those with a QCNN hidden layer support only Estimator.}
\label{table: NNArc}
\centering
\begin{tabular}{lllc}
\hline\hline
\multicolumn{4}{c}{\textbf{(a) CNN Architectures}} \\ \hline
Engine & Layer & Kernel Size & Output Size \\ \hline
\multirow{4}{*}{Stronger} & Conv & $3 \times 3$ & $1 \times 1 \times 64$ \\
& Flatten & - & $64$ \\
& Fully Connected 1 & - & $128$ \\
& Fully Connected 2 & - & $9$ \\
\hline
\multirow{3}{*}{Weaker} & Conv & $3 \times 3$ & $1 \times 1 \times 16$ \\
& Flatten & - & $16$ \\
& Fully Connected & - & $9$ \\
\hline\hline
\multicolumn{4}{c}{\textbf{(b) QNN Architectures}} \\
\hline
& Layer & & Output Size \\ \hline
& Flatten & & $9$ \\
& QNN & & $9$ \\
\hline\hline
\multicolumn{4}{c}{\textbf{(c) QCNN Architectures}} \\ \hline
& Layer & & Output Size \\ \hline
 & Flatten &  & $18$ \\
 & QCNN &  & $9$ \\
\hline\hline
\multicolumn{4}{c}{\textbf{(d) HNN Architectures}} \\ \hline
Output method & Layer &  & Output Size \\ \hline
\multirow{3}{*}{Sampler} & Fully Connected 1 &  & $N$ \\
& QNN &  & $2^N$ \\
& Fully Connected 2 &  & $9$ \\ \hline
\multirow{3}{*}{Estimator} & Fully Connected 1 &  & $N$ \\
& QNN/QCNN &  & $N$ \\
& Fully Connected 2 &  & $9$ \\
\hline\hline
\end{tabular}
\end{table}

These engines used the Tanh function as the activation function before outputting.

We adopt epsilon-greedy decay strategy to train all engines by self-play which is using same engines for battle to update weight parameters. The evaluation was conducted based on round-robin matches, battled for 100 games per one engine. Elo ratings were updated every 100 games from initial rate 1500, using $K=32$ in Eq.~(\ref{eq:elo_update}).

\subsection{Performance Comparison by Architecture}
Table.~\ref{table: engines_result} summarizes Elo ratings of representative models. The models listed in the table are the top-performing examples for each architecture; the full results for all 54 engines are detailed in the Appendix~\ref{sec_engine_result}. In this research, the observed fluctuation in Elo rating for a given model was approximately 20-30 points. As shown in Fig.~\ref{fig: elo}, this level of difference corresponds to a variation in win rate of less than 6\%. We therefore consider a rating difference of approximately 70 or more to be indicative of a significant performance gap.

\begin{table}[htbp]
\begin{center}
\caption{Main engines and the Elo rating: The rating is the final result from a round-robin tournament where it was updated every 100 games.}
\label{table: engines_result}
\begin{tabular}{ccc}
\hline
type & model & rating \\ \hline
Sampler 16 qubits & TPE+RealAmplitudes & 1624.44\\
Sampler 8 qubits & TPE+RealAmplitudes & 1598.82\\
Estimator 16 qubits & HEE+RealAmplitudes & 1578.74\\
Estimator 8 qubits & \begin{tabular}{c}ZZFeatureMap\\+RealAmplitudes\end{tabular} & 1554.54\\
Classical neural network & stronger & 1546.19\\
Estimator 18 qubits (only QCNN) & HEE+QCNN & 1468.43\\
Estimator 9 qubits (only QNN) & TPE+EfficientSU2 & 1455.61\\
\hline
\end{tabular}
\end{center}
\end{table}

The two CNN engines listed give a rough idea of the position of the HNN engines.
HNN with QNN engines achieved the highest rating of 1624.44, followed by classical CNN engines, then QCNN engines, and QNN engines. We can see that ratings are higher as the number of qubits is increased in these settings. These engines are picked up from different categories, and they differ in the combinations of the circuits, which is the highest.

This rating result can be considered a single example. The ratings for the classical and HNN engines are likely influenced by the construction of the classical layer, the methods of encoding classical information (like the board state) into quantum states, and the process of extracting results via measurement. In this research, we employed general-purpose quantum circuits; therefore, there is likely room for performance improvement by implementing architectures more specialized for tic-tac-toe.

Although the Sampler 16 qubit and Sampler 8 qubit engines achieved the highest ratings, it is believed that their classical layers are a significant contributing factor. In the Sampler method, the classical layer following the quantum layer has a remarkably large number of parameters, which can be thought to increase the engine's strength. On the other hand, the Estimator 16 qubits and Estimator 8 qubits engines, which have fewer parameters, still recorded higher ratings than the classical neural network. This suggests that the number of parameters is not necessarily the sole determinant of an engine's strength.

The engines that used only a quantum layer had lower ratings than the classical engines. This indicates that there is room for improvement in areas such as inputting the board state information and processing the output from the neural network. As mentioned earlier, it also highlights the potential for enhancement by implementing quantum circuits more specialized for tic-tac-toe.

\begin{figure}[htbp]
\centering
\includegraphics[width=0.9\linewidth]{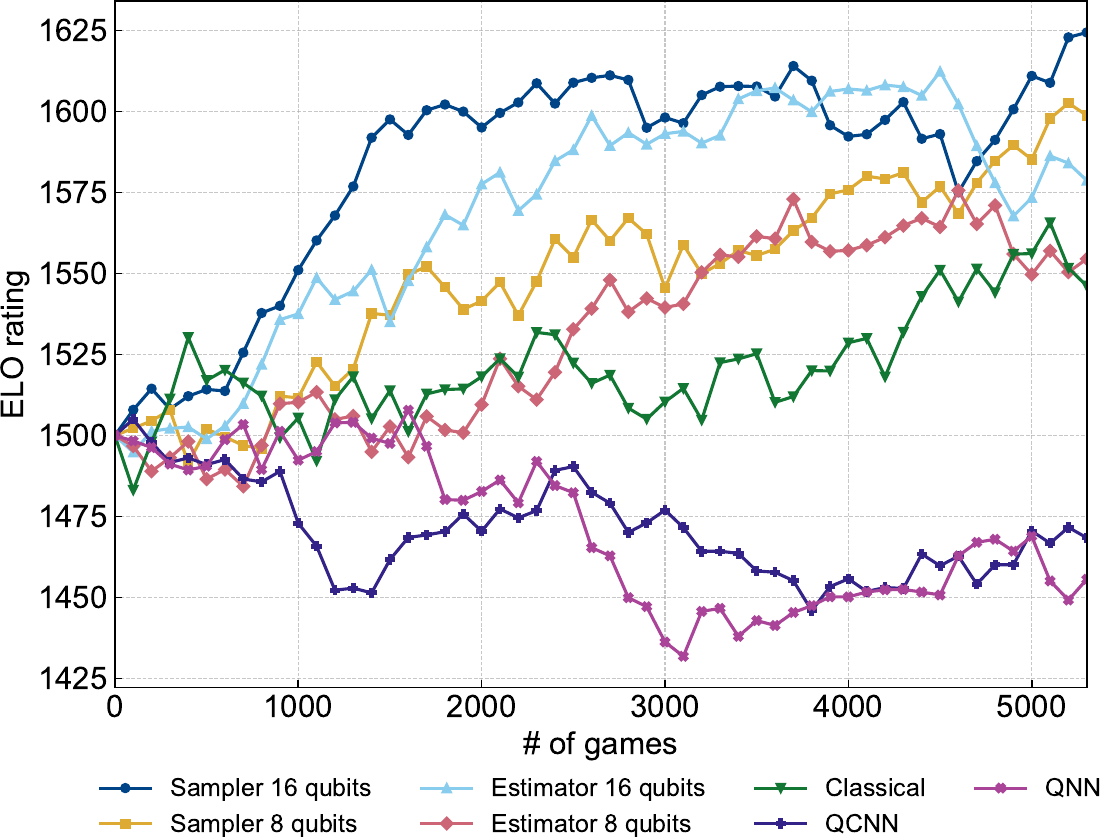}
\caption{Rating progression of the main engines: Since each engine plays 100 games, ratings are calculated every 100 games. Each curve corresponds to the engines in Table.~\ref{table: engines_result}.}
\label{fig: rating}
\end{figure}

Fig.~\ref{fig: rating} shows the rating progression of the main engines. We can see the engines appear to be broadly divided into two groups: the top five engines, including the classical neural network, and the two engines constructed only with quantum layers. The higher-ranked engines tend to increase their ratings in a relatively stable manner, while the lower-ranked engines show a tendency to stagnate around their initial rating values. Since the rating calculation is based on the number of wins and losses, excluding draws, it can be inferred that a stagnating rating implies a higher number of draws.

\subsection{Demonstration Using IBM Quantum Hardware}
\label{sec_demo}
To validate the feasibility of executing the proposed engine on real quantum hardware, we deployed the trained model the strongest in Table.~\ref{table: engines_result} on the \texttt{ibm\_torino} backend via the IBM Quantum platform. Two matches were conducted: one where the quantum engine played first and another where it played second.

\begin{figure}[htbp]
\centering
\includegraphics[width=\linewidth]{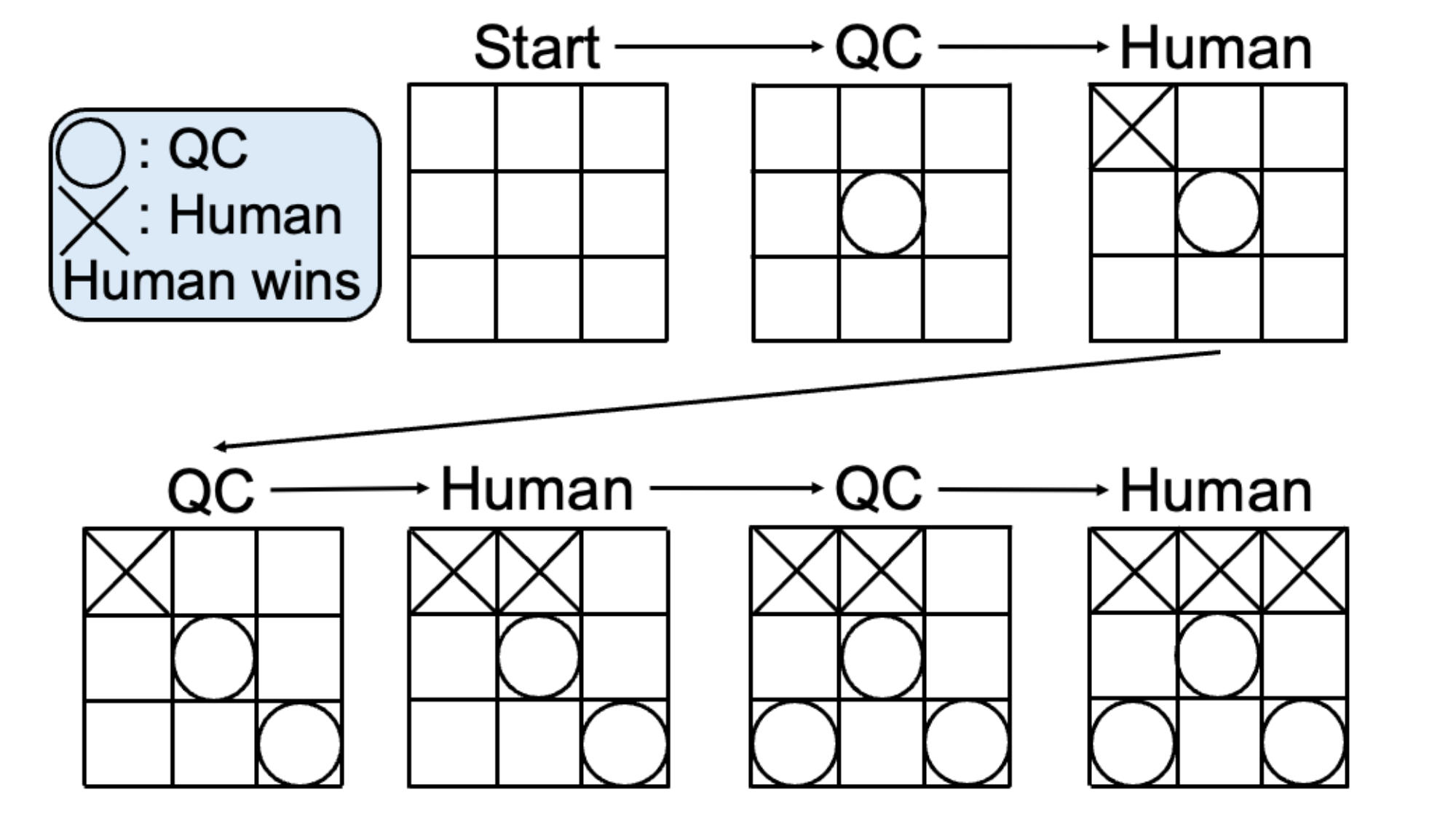}
\caption{Quantum computer vs Human: The quantum computer was first and a human was second.}
\label{fig:qc_vs_human}
\end{figure}

In Fig.~\ref{fig:qc_vs_human}, the quantum computer was first. The first move was good, but in the middle of the game, it could be seen that it was trying to align diagonally. However, the game was over because, in addition to trying to form a line in the area human already put, without defense when human tried to align horizontally in the endgame.

\begin{figure}[htbp]
\centering
\includegraphics[width=\linewidth]{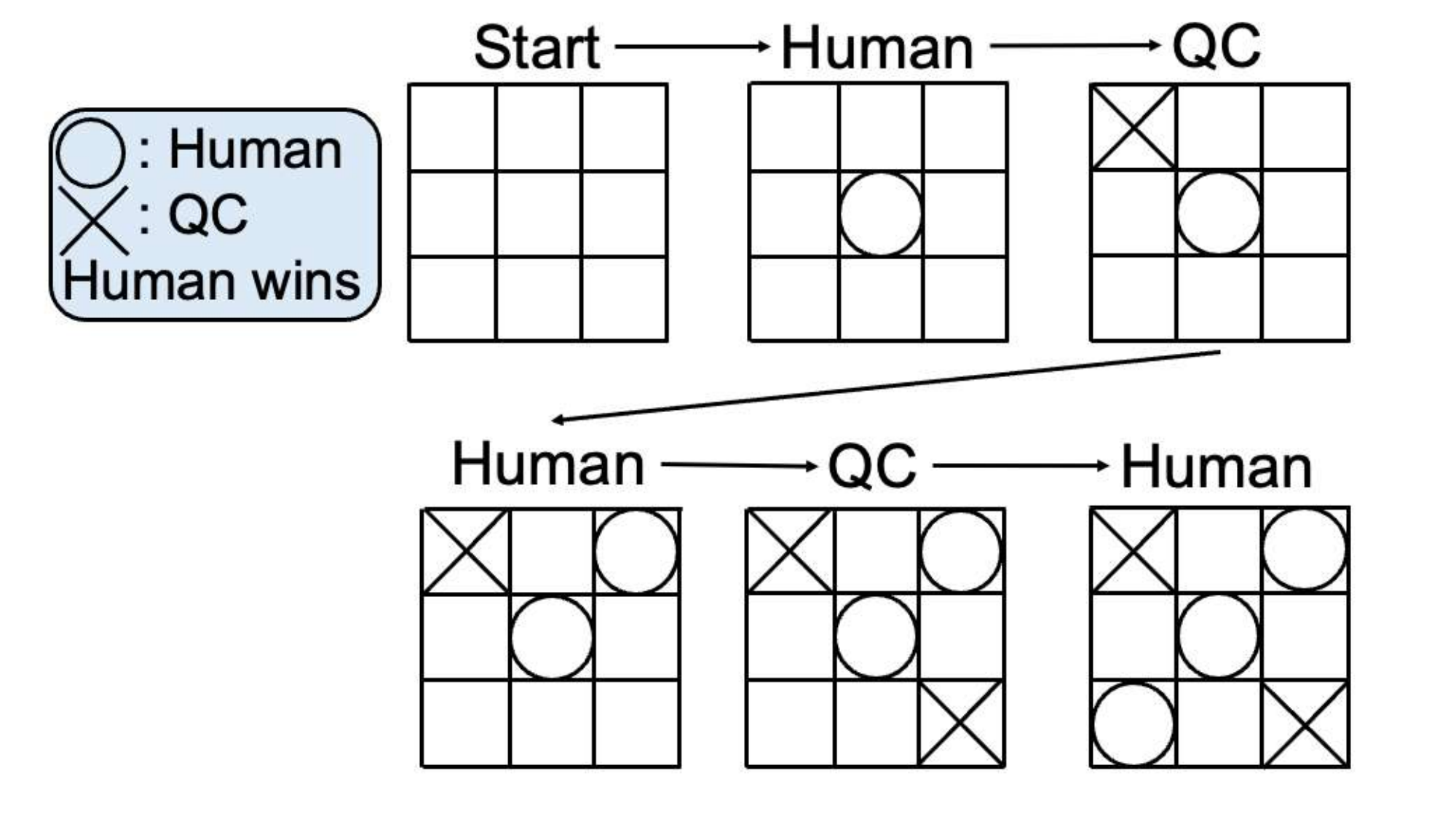}
\caption{Human vs Quantum computer: A human was first and the quantum computer was second.}
\label{fig:human_vs_qc}
\end{figure}

In Fig.~\ref{fig:human_vs_qc}, the quantum computer was second. The quantum computer's move showed that it was trying to align diagonally, but it did not interfere with the human move and the game was over.

As shown in Fig.~\ref{fig:qc_vs_human} and Fig.~\ref{fig:human_vs_qc}, the quantum engine demonstrated an ability to align symbols strategically, but failed to defend effectively in the endgame. This indicates that while the model captures some tactical features, further training or task-specific tuning may be necessary for robustness.

These results demonstrate that even with current hardware constraints, quantum-based game engines can be used in real-world interactive settings.

\section{The Overhead of Engines through QI Model}
\label{sec_qimodel}
Building upon the client-server quantum machine learning framework described in Section~\ref{sub_q-comm}, we evaluate the performance degradation caused by noisy quantum channels in two representative models (Model 1 and Model 2).
\subsection{Two Toy Model Settings}

\begin{figure*}[htbp]
\begin{center}
\includegraphics[width=0.7\linewidth]{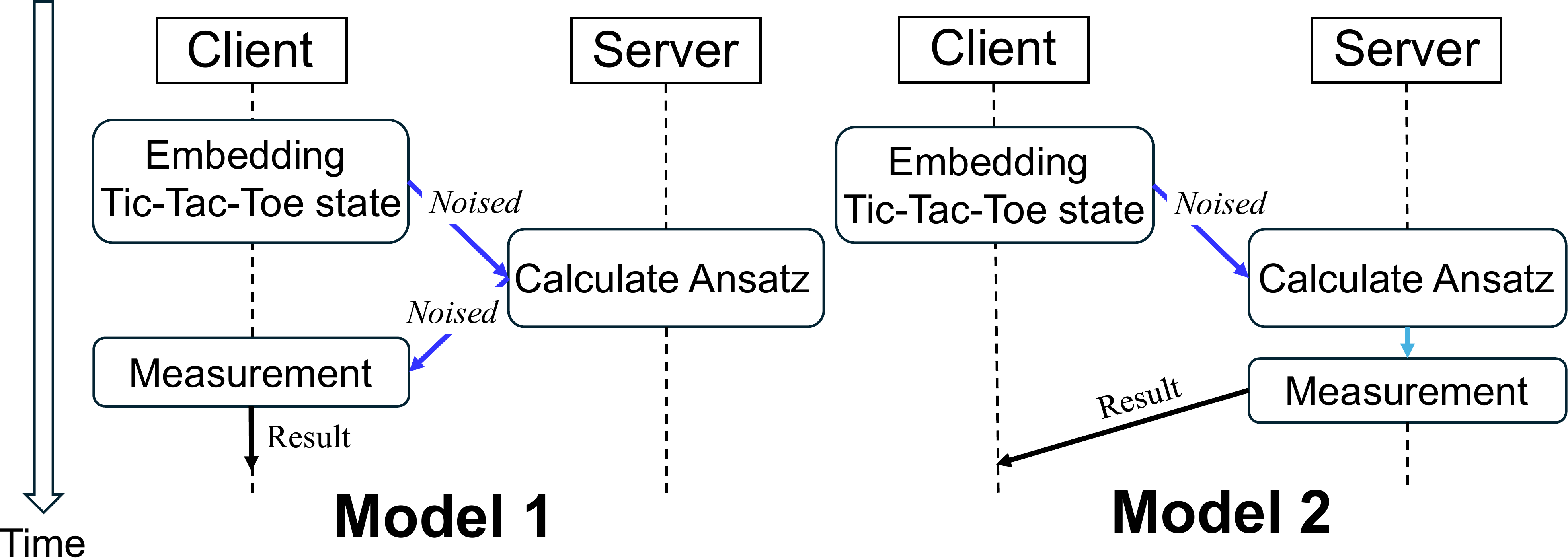}
\caption{Quantum communication model: In Model 1, the client embeds classical data into a qubit, passes it to the server, which calculates the ansatz and then returns it to the client. The client measures the qubit and obtains an output.
In Model 2, the client embeds classical data into qubits, passes them to the server, calculates the ansatz, and then makes a measurement. The measurement outcomes are returned to the client and an output is obtained.}
\label{fig: model}
\end{center}
\end{figure*}

We consider two models of quantum communication channels, depending on the quantum resources availability on the client side.  

\begin{itemize}
\item \textbf{Model 1}: The client can embed quantum states and perform measurements.  
\item \textbf{Model 2}: The client cannot measure the quantum states after server-side computation; instead, the server performs the measurement and returns the outcomes to the client.
\end{itemize}

A schematic diagram of these quantum communication models is shown in Fig.~\ref{fig: model}. In both models, only noise in the quantum communication channel is considered.  

We employ a 8-qubit HNN using Estimator, with different Embedding. Noise is introduced by inserting rotation gates between the quantum circuits of the engine, with its parameters randomized to pseudo-realize the noise effect.

The noise strength depends on the quantum communication distance and is modeled by the following function:
\begin{align}
\sigma(d) = 10^{\frac{\alpha}{10}d} - 1,
\label{math:noise_distance}
\end{align}

where $\sigma(d)$ represents the standard deviation of the noise strength, and $\alpha = 0.2\, \mathrm{dB/km}$. Gaussian noise is applied to each parameter with a standard deviation of $\sigma(d)$.  

The evaluation consists of two experiments:  

\begin{enumerate}
\item Fixed Distance Experiment: We assess the engine performance under three noise patterns, with the distance set to $100\, \mathrm{km}$.  
\begin{itemize}
\item \textbf{Pattern A}: Noiseless during training and evaluation.  
\item \textbf{Pattern B}: Noiseless during training, but noisy during evaluation.  
\item \textbf{Pattern C}: Noised during both training and evaluation.  
\end{itemize}
\item Distance-Dependent Experiment: We vary the communication distance from $10^{-2}\, \mathrm{km}$ to $10\, \mathrm{km}$ and evaluate the engine under Pattern B.
\end{enumerate}

In this section, we determine the final rating by playing 10,000 games against a random opponent and updating the rating every 100 games.

\subsection{The Effect of Noisy Quantum Communication}
Table.~\ref{table: qi} presents the results for different noise models and engines. 

When comparing Pattern A and Pattern B, almost all engines achieved higher ratings with Pattern A than with Pattern B. This indicates that the ratings are susceptible to noise. However, the extent of the rating drop varies depending on the engine.

When comparing Pattern B and Pattern C, some of the engines show that Pattern B yields higher ratings, and others show that Pattern C's rating is higher or there is almost no change. This suggests that while noise can adversely affect an engine's performance, in some cases, it can conversely enhance it. We believe this is likely influenced by factors such as the engine's level of training, for example, the size of the feature space it explores.

The degree of rating fluctuation due to the patterns varies for each engine. While some engines show little difference depending on the pattern, others exhibit moderate fluctuations. Those that fluctuate significantly are thought to be heavily influenced by noise. We believe this is also heavily influenced by the way the quantum circuits are constructed.

\begin{table}[htbp]
\caption{The result of using different noise models and engines: The table features three different engines in each model, and the rating is updated every 100 games and finally determined by playing through 10,000 games against a random engine.}
\label{table: qi}
\begin{center}
\begin{tabular}{cccl}
\hline
engine & model & noise pattern & rating\\ \hline
\multirow{6}{*}{\shortstack{Estimator \\ ZFeatureMap \\ RealAmplitudes}} & \multirow{3}{*}{1} & A & 1515.26 \\
&& B & 1494.63 \\
&& C & 1535.15 \\ \cline{2-4}
 & \multirow{3}{*}{2} & A & 1523.31 \\
&& B & 1497.44 \\
&& C & 1494.44 \\ \hline
\multirow{6}{*}{\shortstack{Estimator \\ ZZFeatureMap \\ RealAmplitudes}} & \multirow{3}{*}{1} & A & 1533.80 \\
&& B & 1547.76 \\
&& C & 1487.83 \\ \cline{2-4}
 & \multirow{3}{*}{2} & A & 1540.22 \\
&& B & 1522.41 \\
&& C & 1528.41 \\ \hline
\multirow{6}{*}{\shortstack{Estimator \\ TPE \\ RealAmplitudes}} & \multirow{3}{*}{1} & A & 1573.57 \\
&& B & 1536.29 \\
&& C & 1510.10 \\ \cline{2-4}
 & \multirow{3}{*}{2} & A & 1526.79 \\
 && B & 1524.69 \\
 && C & 1527.20 \\
 \hline
\end{tabular}
\end{center}
\end{table}

\begin{figure}[htbp]
\centering
\includegraphics[width=0.9\linewidth]{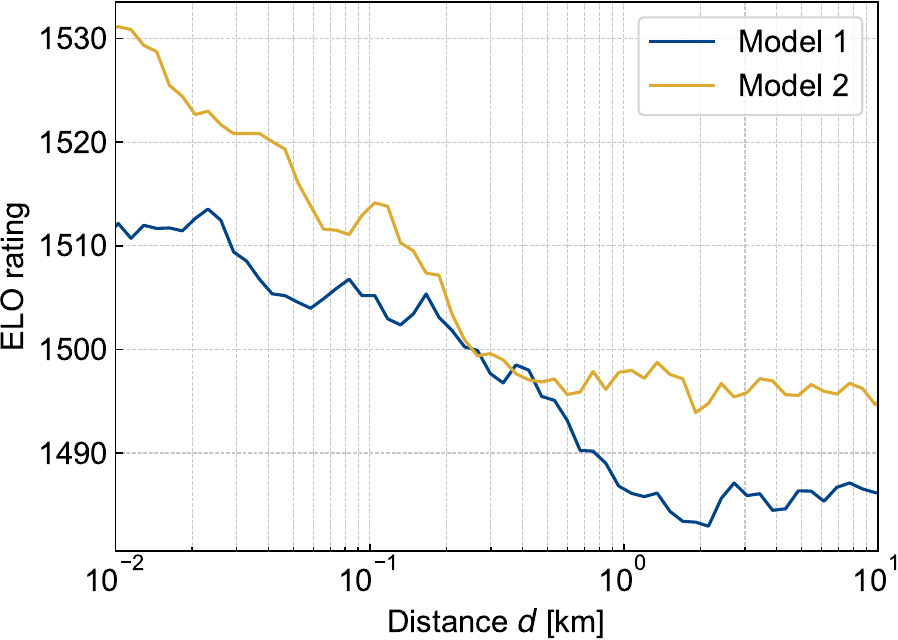}
\caption{Transition of ratings per distance: The type of engine used in this graph is the following; embedding is \texttt{HEE}, the ansatz is \texttt{RealAmplitudes}, and the noise pattern is Pattern B.}
\label{fig: he_per_noise}
\end{figure}

Fig.~\ref{fig: he_per_noise} illustrates the rating transition as a function of distance. As the figure shows, the rating gradually decreases as the distance $d$ increases. For the engines used in this research, when comparing Model 1 and Model 2, the rate of this decrease appears to be roughly the same. This indicates that the magnitude of the noise affects the rating. Then, around $d = 10$, the rating tends to level off, indicating that the engine's sensitivity to noise saturates at longer distances. When comparing Model 1 and Model 2, the rating for Model 2 begins to stagnate at a smaller distance. This can be attributed to the fact that the magnitude of noise affecting Model 2 is half that of Model 1, and therefore, the impact on its rating is also smaller.

\begin{figure*}[ht]
\begin{center}
\includegraphics[width=\linewidth]{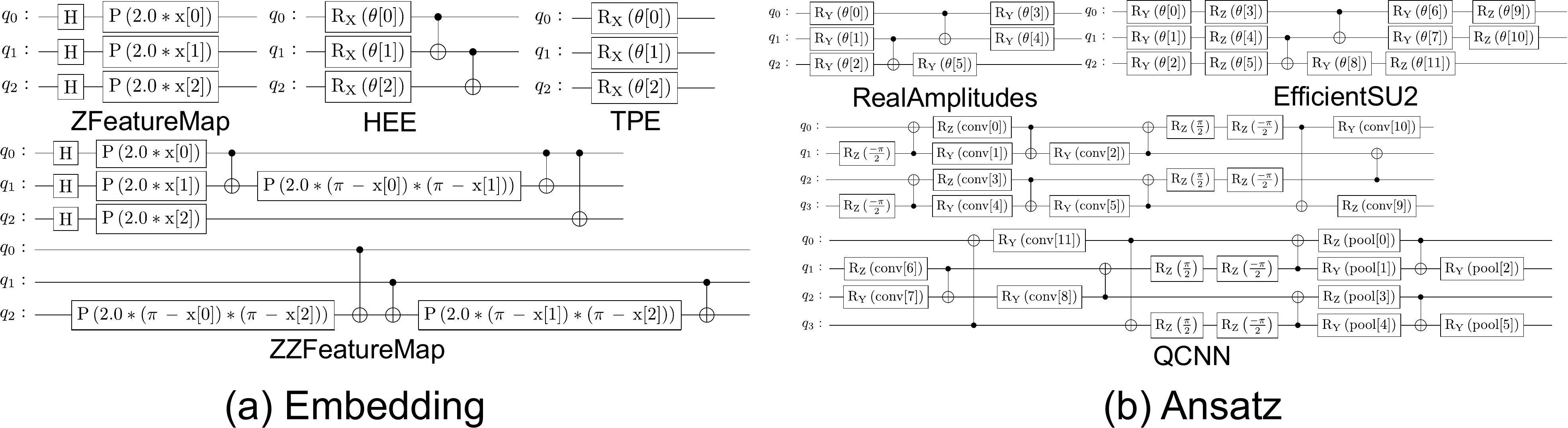}
\caption{Embedding and Ansatz used in engines: (a) Embedding is a quantum circuit for embedding certain classical data into qubits. \texttt{ZFeatureMap} and \texttt{ZZFeatureMap} were built using the qiskit library, while \texttt{HEE} and \texttt{TPE} were constructed by referring to the circuit for embedding in ~\cite{Thanasilp_2023}. (b) Ansatz is a circuit for processing data embedded in qubits using weight parameters. \texttt{RealAmplitudes} and \texttt{EfficientSU2} were built using the qiskit library. QCNN referred to the circuit in~\cite{Vatan_2004} and Qiskit machine learning tutorial.}
\label{fig: embed_and_ansatz}
\end{center}
\end{figure*}

\section{Conclusion}
\label{sec_conclusion}
In this research, we built game engines using classical, quantum, and classical-quantum hybrid neural networks and compared their performances. 
We introduced Elo rating to comprehensively evaluate classical and quantum systems, and by comparing the performance of each engine in terms of ratings, we can determine the relative strength of the engines.
In the same way that the development of computers and algorithms has led to the development of classical game engines such as chess, the development of quantum computers and quantum algorithms is expected to contribute to the development of quantum game engines as they are developed.

In this paper, the classical-quantum hybrid method achieved better results than the classical method. Although this result is expected to vary depending on how each engine is constructed and the type of game being played, the objective of comparing the different engines through the rating was achieved.
We also modeled a simple noisy communication channel, which can be used not only for the game solver task but also as a guide for developing various quantum network applications, and verified the amount of overhead observed. 
This is a necessary indicator not only for simple applications but also for the development of hidden applications.
It will be a stepping stone to develop use cases of quantum network applications that are better than classical ones, while taking noise into account.

Building on this foundation, there are several promising avenues for future research. It would be valuable to develop more comprehensive frameworks that establish unified standards for comparing quantum and classical systems. This includes exploring more complex tasks where neither paradigm holds a clear, pre-existing advantage, thereby providing a more neutral ground for performance evaluation. Furthermore, as quantum hardware continues to scale, investigating these larger competitive environments and refining the corresponding evaluation methodologies will be crucial in the ongoing search for quantum advantage.

\appendices
\section{Quantum Circuits used in the engines}
\label{app_qc}

Fig.~\ref{fig: embed_and_ansatz} shows the Embedding and Ansatz used in this paper.
\texttt{ZFeatureMap}, \texttt{ZZFeatureMap}, \texttt{RealAmplitudes} and \texttt{EfficientSU2} are reffered to the qiskit library. 
\texttt{HEE} (Hardware Efficient Embedding) and \texttt{TPE} (Tensor Product Embedding) are reffered to the circuit for embedding in ~\cite{Thanasilp_2023}, and QCNN is reffered to the circuit in ~\cite{Vatan_2004} and Qiskit machine learning.

\section{Engine Properties and Rating}
\label{sec_engine_result}
Table~\ref{table: rating engines} lists the ratings and detailed properties of all 46 engines constructed in this research. The engines are broadly divided into three types, classical, classical-quantum hybrid, and quantum. Within those categories, they are further subdivided by qubit count, post-measurement processing, and the choice of Embedding and Ansatz methods.

First, for the classical CNN engines (row 1), we created two versions, "Stronger" and "Weaker". The table shows that the difference in their ratings can be attributed to the large difference in their parameter counts.

Next, for the Hybrid NN engines (rows 2-5), their respective ratings do not change significantly with the number of classical parameters. This suggests that the quantum circuit configuration is the dominant contributor to performance, particularly the choice of Embedding and Ansatz. Furthermore, for the tic-tac-toe task, an increase in circuit depth tends to lower the rating. This is likely task-dependent; for a simple task like tic-tac-toe, a shallower circuit may be sufficient, and a larger depth could lead to training difficulties within our experimental setup.

Finally, the QNN and QCNN engines show low overall ratings, which resulted in a rating below the initial 1500. This may be due to issues with the compatibility of this specific Embedding and Ansatz pairing, which warrants further investigation.
\begin{table*}[htbp]
\caption{Properties and Elo ratings of 54 engines: This table is divided into seven rows, which can be classified as follows: (a) the first row, (b) rows 2-5, and (c) rows 6-7. (a) is 2 models of engines using only CNN. (b) is 40 models of engines using HNN with QNN or QCNN. (c) is 12 models of engines using only QNN or QCNN. The rating is the final result from a round-robin tournament where it was updated every 100 games, and the number of parameters indicates the number of trainable weights for the classical and quantum layers, respectively. The number of CX gates and the quantum circuit depth are shown for reference.}
\label{table: rating engines}
\begin{center}
\begin{tabular}{ll|c|c|c|c|cc}
\hline
type & model & rating & classical params & quantum params & total params & CX gates & depth\\ 
\hline
\multirow{2}{*}{classical neural network} & stronger & \textbf{1546.19} & 10057 & - & 10057 & - & -\\ 
 & weaker & 1388.85 & 297 & - & 297 & - & -\\ \hline \hline
\multirow{8}{*}{Sampler 8 qubits} & ZFeatureMap+RealAmplitudes & 1418.76 & \multirow{8}{*}{2393} & \multirow{4}{*}{16} & \multirow{4}{*}{2409} & 7 & 11\\ 
 & ZZFeatureMap+RealAmplitudes & 1559.11 &  &  &  & 63 & 50\\ 
 & HEE+RealAmplitudes & 1587.29 &  &  &  & 14 & 17\\ 
 & TPE+RealAmplitudes & \textbf{1598.82} &  &  &  & 7 & 10\\ \cline{5-6}
 & ZFeatureMap+EfficientSU2 & 1584.18 &  & \multirow{4}{*}{32} & \multirow{4}{*}{2425} & 7 & 13\\ 
 & ZZFeatureMap+EfficientSU2 & 1575.87 &  &  &  & 63 & 52\\ 
 & HEE+EfficientSU2 & 1556.22 &  &  &  & 14 & 19\\ 
 & TPE+EfficientSU2 & 1580.00 &  &  &  & 7 & 12\\ \hline
\multirow{8}{*}{Sampler 16 qubits} & ZFeatureMap+RealAmplitudes & 1579.19 & \multirow{8}{*}{589913} & \multirow{4}{*}{32} & \multirow{4}{*}{589945} & 15 & 19\\ 
 & ZZFeatureMap+RealAmplitudes & 1487.97 &  &  &  & 255 & 106\\ 
 & HEE+RealAmplitudes & 1607.00 &  &  &  & 30 & 33\\ 
 & TPE+RealAmplitudes & \textbf{1624.44} &  &  &  & 15 & 18\\ \cline{5-6}
 & ZFeatureMap+EfficientSU2 & 1562.35 &  & \multirow{4}{*}{64} & \multirow{4}{*}{589977} & 15 & 21\\ 
 & ZZFeatureMap+EfficientSU2 & 1552.84 &  &  &  & 255 & 108\\ 
 & HEE+EfficientSU2 & 1623.85 &  &  &  & 30 & 35\\ 
 & TPE+EfficientSU2 & 1569.38 &  &  &  & 15 & 20\\ \hline
\multirow{12}{*}{Estimator 8 qubits} & ZFeatureMap+RealAmplitudes & 1537.84 & \multirow{8}{*}{161} & \multirow{4}{*}{16} & \multirow{4}{*}{177} & 7 & 11\\ 
 & ZZFeatureMap+RealAmplitudes & \textbf{1554.54} &  &  &  & 63 & 50\\ 
 & HEE+RealAmplitudes & 1553.05 &  &  &  & 14 & 17\\ 
 & TPE+RealAmplitudes & 1416.87 &  &  &  & 7 & 10\\ \cline{5-6} 
 & ZFeatureMap+EfficientSU2 & 1493.48 &  & \multirow{4}{*}{32} & \multirow{4}{*}{193} & 7 & 13\\ 
 & ZZFeatureMap+EfficientSU2 & 1543.56 &  &  &  & 63 & 52\\ 
 & HEE+EfficientSU2 & 1494.05 &  &  &  & 14 & 19\\ 
 & TPE+EfficientSU2 & 1489.77 &  &  &  & 7 & 12\\ \cline{4-6}
 & ZFeatureMap+QCNN & 1521.20 & \multirow{4}{*}{125} & \multirow{4}{*}{36} & \multirow{4}{*}{161} & 32 & 17\\
 & ZZFeatureMap+QCNN & 1527.72 & & & & 88 & 56\\
 & HEE+QCNN & 1552.64 & & & & 39 & 23\\
 & TPE+QCNN & 1491.45 & & & & 32 & 16\\ \hline
\multirow{12}{*}{Estimator 16 qubits} & ZFeatureMap+RealAmplitudes & 1487.63 & \multirow{8}{*}{313} & \multirow{4}{*}{32} & \multirow{4}{*}{345} & 15 & 19\\ 
 & ZZFeatureMap+RealAmplitudes & 1506.73 &  &  &  & 255 & 106\\ 
 & HEE+RealAmplitudes & \textbf{1578.74} &  &  &  & 30 & 33\\ 
 & TPE+RealAmplitudes & 1503.81 &  &  &  & 15 & 18\\ \cline{5-6}
 & ZFeatureMap+EfficientSU2 & 1532.11 &  & \multirow{4}{*}{64} & \multirow{4}{*}{377} & 15 & 21\\ 
 & ZZFeatureMap+EfficientSU2 & 1547.77 &  &  &  & 255 & 108\\ 
 & HEE+EfficientSU2 & 1476.19 &  &  &  & 30 & 35\\ 
 & TPE+EfficientSU2 & 1495.04 &  &  &  & 15 & 20\\ \cline{4-6}
 & ZFeatureMap+QCNN & 1453.57 & \multirow{4}{*}{241} & \multirow{4}{*}{72} & \multirow{4}{*}{313} & 64 & 17\\
 & ZZFeatureMap+QCNN & 1482.25 & & & & 304 & 104\\
 & HEE+QCNN & 1496.90 & & & & 79 & 31\\
 & TPE+QCNN & 1518.80 & & & & 64 & 16\\ \hline \hline
\multirow{8}{*}{\begin{tabular}{c}Estimator 9 qubits \\ (only QNN)\end{tabular}} & ZFeatureMap+RealAmplitudes & 1391.46 & - & \multirow{4}{*}{18} & \multirow{4}{*}{18} & 8 & 12\\ 
 & ZZFeatureMap+RealAmplitudes & 1374.80 & - &  &  & 80 & 57\\ 
 & HEE+RealAmplitudes & 1319.18 & - &  &  & 16 & 19\\ 
 & TPE+RealAmplitudes & 1313.40 & - &  &  & 8 & 11\\ \cline{5-6}
 & ZFeatureMap+EfficientSU2 & 1348.71 & - & \multirow{4}{*}{36} & \multirow{4}{*}{36} & 8 & 14\\
 & ZZFeatureMap+EfficientSU2 & 1425.78 & - & & & 80 & 59\\
 & HEE+EfficientSU2 & 1393.57 & - & & & 16 & 21\\
 & TPE+EfficientSU2 & \textbf{1455.61} & - & & & 8 & 13\\ \hline
\multirow{4}{*}{\begin{tabular}{c}Estimator 18 qubits \\ (only QCNN)\end{tabular}} & ZFeatureMap+QCNN & 1419.14 & - & \multirow{4}{*}{81} & \multirow{4}{*}{81} & 72 & 17\\ 
 & ZZFeatureMap+QCNN & 1423.51 & - &  &  & 378 & 116\\ 
 & HEE+QCNN & \textbf{1468.43} & - &  &  & 89 & 33\\ 
 & TPE+QCNN & 1408.37 & - &  &  & 72 & 16\\
\hline
\end{tabular}
\end{center}
\end{table*}

\bibliography{kame.bib}
\bibliographystyle{unsrt}

\EOD
\end{document}